\documentclass[11pt]{article}
\usepackage{mathrsfs}
\usepackage[centertags]{amsmath}
\usepackage{amsfonts}
\usepackage{amssymb}
\usepackage{amsthm}
\usepackage{cases}
\usepackage{indentfirst}
\usepackage{epsfig}
\usepackage {graphicx}
\usepackage{color}
\usepackage{CJK}

\date{}

\newtheorem{theorem}{Theorem}[section]
\newtheorem{corollary}{Corollary}[section]
\newtheorem{lemma}{Lemma}[section]

\newtheorem{remark}{Remark}[section]
\newtheorem{definition}{Definition}[section]
\numberwithin{equation}{section}

\allowdisplaybreaks[4]

\begin{document}
\title{\textbf{A sharp sufficient condition of block signal recovery via $l_2/l_1$-minimization}\thanks{
{Corresponding author. E-mail address: wjj@swu.edu.cn (J. Wang).} }}
\author{Jianwen Huang$^a$,\quad Jianjun Wang$^a$,\quad Wendong Wang$^b$,\quad Feng Zhang$^a$\\
{\small $^a$School of Mathematics and Statistics, Southwest
University, Chongqing, 400715, China}\\{\small $^b$School of Computer and Information Science, Southwest
University, Chongqing, 400715, China}\\{\small}}

\maketitle
\begin{quote}
{\bf Abstract.}~~This work gains a sharp sufficient condition on the block restricted isometry property for the recovery of sparse signal. Under the certain assumption, the signal with block structure can be stably recovered in the present of noisy case and the block sparse signal can be exactly reconstructed in the noise-free case. Besides, an example is proposed to exhibit the condition is sharp. As byproduct, when $t=1$, the result improves the bound of block restricted isometry constant $\delta_{s|\mathcal{I}}$ in Lin and Li (Acta Math. Sin. Engl. Ser. 29(7): 1401-1412, 2013).

\end{quote}

{\bf AMS Classification(2010)}:~~94A12, 94A15, 94A08, 68P30, 41A64, 15A52, 42C15

\section{Introduction}
\label{sec1}

Compressed sensing is one of novel sampling theory, recently attracting more and more researchers' interest (see e.g. \cite{Bryant and Cathain,Cahill et al,Fang et al,Kong et al,Wang et al 2015,Wen et al,Xu et al,Zeng et al 2012,Zeng et al 2014}). It plays a critical role in a variety
 of fields such as signal processing, machine learning, seismology, electrical engineering and statistics. In compressed sensing, we are interested in
 recovering an unknown signal $x\in \mathbb{R}^N$ that fulfils the undetermined system of linear equations, that is,
\begin{align}\label{eq.1}
b=\Phi x+\xi
\end{align}
where $\Phi\in \mathbb{R}^{M\times N}$ is a known sensing matrix with $M\ll N$, observed signal $b\in \mathbb{R}^M$ and $\xi\in \mathbb{R}^M$ is an unknown bounded noise. In particular, when the noise vector $\xi=0$, the linear measurement (\ref{eq.1}) reduces to the noiseless situation, namely,
\begin{align}\label{eq.2}
b=\Phi x.
\end{align}
It is well known that there is not only unique solution to the linear measurement (\ref{eq.1}). However, we assume that the signal $x$ consists of a small number of nonzero coefficients that spread arbitrarily throughout the signal, that is, suppose that $x$ is sparse. Under this assumption, the problem has a unique sparse solution. Initially, the way that solves it is to study the $l_0$-minimization, i.e.,
\begin{align}\label{eq.3}
\min_{x} \|x\|_0,~\mbox{subject to}~b=\Phi x,
\end{align}
where $\|x\|_0$ counts the number of nonzero elements of the vector $x$. However, it is nonconvex and NP-hard and accordingly is not to solve efficiently in the polynomial time. It is now well understood that the $l_1$-minimization approach offers an effective method for resolving this problem, i.e.,
\begin{align}\label{eq.4}
\min_{x} \|x\|_1,~\mbox{subject to}~b=\Phi x,
\end{align}
where $\|x\|_1=\Sigma^N_{i=1}|x_i|$. Formally, the $l_1$-minimization (\ref{eq.4}) is convex and therefore is computationally tractable. The equivalency \cite{Candes} between the problem (\ref{eq.3}) and the issue (\ref{eq.4}) has been proved by making use of the restricted isometry property (RIP) with some restricted isometry constant (RIC). Let $s$ is a positive integer with $1\leq s\leq N$, the restricted isometry constant $\delta_s$ of order $s$ of a matrix $\Phi$ is defined as the smallest nonnegative constant such that
\begin{align}\label{eq.5}
1-\delta_s\leq\frac{\|\Phi x\|^2_2}{\|x\|^2_2}\leq1+\delta_s
\end{align}
holds for any $s$-sparse vectors $x\in \mathbb{R}^N$. Here, we say that $x\in \mathbb{R}^N$ is $s$-sparse that $\|x\|_0\leq s\ll N$.

However, in a lot of practical applications, some real-world signals may exhibit some particular sparsity patterns, where the non-zero coefficients arise in some fixed blocks. These non-conventional signals have a number of potential applications in various fields of science and technology, like DNA
microarrays \cite{Parvaresh et al}, equalization of sparse communication channels \cite{Cotter and Rao}, face recognition \cite{Elhamifar and Vidal}, source localization \cite{Malioutov et al}, reconstruction of multi-band signals \cite{Simila and Tikka} and
multiple measurement vector model \cite{Eldar and Rauhut}. We refer to these signals as block sparse signals. Literature \cite{Eldar and Mishali} first introduced the concept of block sparsity. Recently, block sparsity recovery has attracted considerable interests; for more details, see \cite{Lin and Li}, \cite{Wang et al 2013}, \cite{Wang et al 2014} and \cite{Gao and Ma}.

We assume that a block sparse signal $x\in\mathbb{R}^N$ over block index set $\mathcal{I}=\{d_1,d_2,\cdots,d_l\}$ can be represented as:
\begin{align}\label{eq.38}
x=[\underbrace{x_1,\cdots,x_{d_1}}_{x[1]},\underbrace{x_{d_1+1},\cdots,x_{d_1+d_2}}_{x[2]},\cdots,\underbrace{x_{N-d_l+1},\cdots,x_N}_{x[l]}]^T,
\end{align}
where $x[i]$ stands for the $i$th block of $x$ associated with the block length $d_i$ and $N=d_1+d_2+\cdots+d_l.$ We say a vector $x\in \mathbb{R}^N$ as block $s$-sparse over index set $\mathcal{I}=\{d_1,\cdots,d_l\}$ when $x[i]$ is non-zero for no more than $s$ indices $i$. In order to reconstruct a block sparse signal, analogous to the $l_0$-minimization, we search for the sparsest block vector by employing the $l_2/l_0$-minimization below proposed by \cite{Elhamifar and Vidal}:
\begin{align}\label{eq.6}
\min_x \|x\|_{2,0},~\mbox{subject to}~b=\Phi x,
\end{align}
where $\|x\|_{2,0}=\sum^l_{i=1}I(\|x[i]\|_2>0)$, and $I(x)$ denotes an indicator function that $I(x)=1$ or $0$ according as $x>0$ or otherwise. Accordingly, we could define a block $s$-sparse vector $x$ as $\|x\|_{2,0}\leq s.$
However, the $l_2/l_0$-minimization approach remains NP-hard and computationally intractable. Let $\|x\|_{2,\mathcal{I}}=\sum^l_{i=1}\|x[i]\|_2.$ Similar to the case of $l_0$-minimization, one natural ideal is to substitute the $l_2/l_0$-minimization with the $l_2/l_1$-minimization below given by \cite{Eldar et al 2010}:
\begin{align}\label{eq.7}
\min_x \|x\|_{2,\mathcal{I}},~\mbox{subject to}~b=\Phi x.
\end{align}
In order to describe the performance of this approach, the block restricted isometry property (block RIP) was defined by \cite{Eldar and Mishali}.
\begin{definition}
Given a sensing matrix $\Phi$ with size $M\times N$, where $M<N$, one says that the measurement matrix $\Phi$ obeys the block RIP over $\mathcal{I}=\{d_1,\cdots,d_l\}$ with constants $\delta_{s|\mathcal{I}}$ if for every vector $x\in\mathbb{R}^N$ with block s-sparse over $\mathcal{I}$ such that
\begin{align}\label{eq.8}
1-\delta_{s|\mathcal{I}}\leq\frac{\|\Phi x\|^2_2}{\|x\|^2_2}\leq1+\delta_{s|\mathcal{I}}
\end{align}
holds. We say the smallest constant $\delta_{s|\mathcal{I}}$ that fulfils the above inequality (\ref{eq.8}) as the block RIC corresponding with the matrix $\Phi.$
\end{definition}

It is easy to see that the block RIP is an generalization of the standard RIP, but it is a less stringent requirement in comparison with the standard RIP. Eldar et al. \cite{Eldar and Mishali} proved that any block $s$-sparse signal could be exactly recovered via the $l_2/l_1$-minimization as the sensing matrix $\Phi$ meets the block RIP with $\delta_{2s|\mathcal{I}}<\sqrt{2}-1\approx0.4142$. One can improve the block RIP, for example, Lin and Li \cite{Lin and Li} improved the bound to $\delta_{2s|\mathcal{I}}<(77-\sqrt{1337})/82\approx0.4931,$ meanwhile obtained another sufficient condition $\delta_{s|\mathcal{I}}<0.307$. Recently, Gao and ma \cite{Gao and Ma} improved that bound to $\delta_{2s|\mathcal{I}}<4/\sqrt{41}\approx0.6246.$ Up to now, to the best of our knowledge, there is no work that further concentrates on improvement of the block RIC. Improving the bound concerning block RIC $\delta_{s|\mathcal{I}}$ could bring several advantages. First of all, in compressed sensing, it permits more sensing matrices to be utilized; Then, it permits for reconstructing a block sparse signal with more non-zero coefficients under the condition of the identical matrix $\Phi$; In the end, it provides better error estimation in a general issue to reconstruct signals with noise and mismodeling error; for more information, see \cite{Lin and Li}, \cite{Eldar and Mishali}, \cite{Gao and Ma}, \cite{Lin and Li b} and \cite{Mo and Li}. The purpose of this article is to discuss the improvement for the block RIC, and we will investigate the following minimization for the noisy and mismodeling measurement $b=\Phi x+\xi$ satisfying $\|\xi\|_2\leq\rho$:
\begin{align}\label{eq.9}
\min_x \|x\|_{2,\mathcal{I}},~\mbox{subject to}~\|\Phi x-b\|_2\leq\rho.
\end{align}
First, the following theorem is one of our main results that give a sufficient condition of recovery as signal $x$ is not block sparse and the measure is corrupted by the noise. For any $x\in\mathbb{R}^N$, we represent $x_{\max(s)}$ as $x$ with all but the largest $s$ blocks in $l_2$ norm set to zero and $x_{-\max(s)}=x-x_{\max(s)}$. Set $\tilde{t}=\max\{\sqrt{t},t\}$.

\begin{theorem}
We assume that the measurement matrix $\Phi$ with size $M\times N(M<N)$ fulfils for $0<t<4/3$, $ts\geq2$,
\begin{align}\label{eq.10}
\delta_{ts|\mathcal{I}}<\frac{t}{4-t}.
\end{align}
If $x^*$ is a solution to problem (\ref{eq.9}), then we have
\begin{align}
\notag\|x^*-x\|_2\leq& \frac{2\sqrt{2}\rho\sqrt{1+\delta_{ts|\mathcal{I}}}}{t+(t-4)\delta_{ts|\mathcal{I}}}\tilde{t}\\
\label{eq.39}&+\frac{1}{2}\sqrt{\frac{2}{s}}\bigg(\frac{8\delta_{ts|\mathcal{I}}+4\sqrt{(t+(t-4)\delta_{ts|\mathcal{I}})\delta_{ts|\mathcal{I}}}}{t+(t-4)\delta_{ts|\mathcal{I}}}
+1\bigg)\|x_{-\max(s)}\|_{2,\mathcal{I}}.
\end{align}
\end{theorem}
\begin{remark}
The inequality (\ref{eq.39}) provides an error upper bound about the noisy recovery utilizing the $l_2/l_1$-minimization methodology (\ref{eq.9}). Especially, the sparsity degree of the signal $s$ and the noise amplitude $\rho$ can control the recovery accuracy of the $l_2/l_1$-minimization.
\end{remark}

\begin{remark}
Theorem $1.1$ shows that any signals of block pattern contaminated by noise, i.e., (\ref{eq.1}) can be stably recovered via the $l_2/l_1$-minimization approach if the sensing matrix $\Phi$ satisfies the block RIP with a appropriate block RIC. Beside, when signal $x$ is block $s$-sparse, then Theorem $1.1$ ensure the signal $x$ can be robustly constructed in the noisy scenario.
\end{remark}

\begin{remark}
It is known that Lin and Li \cite{Lin and Li} established a sufficient condition $\delta_{s|\mathcal{I}}<0.307$ for robust recovery. In Theorem $1.1$, when $t=1$, its result enhances the bound on the block RIC to $\delta_{s|\mathcal{I}}<1/3\approx 0.3333.$
\end{remark}

\begin{remark}
When the block size $d_i=1~(i=1,\cdots,l)$, the result of Theorem $1.1$ degenerates to the convention case consistent with the results of \cite{Zhang and Li}.
\end{remark}

\begin{remark}
In the proof process of Theorem $1.1$, from the inequality (\ref{eq.37}) and applying Lemma $5.3$ \cite{Cai and Zhang}, then we could another error estimation as follows:
\begin{align}
\notag\|x^*-x\|_2\leq& \frac{2\sqrt{2}\rho\sqrt{1+\delta_{ts|\mathcal{I}}}}{t+(t-4)\delta_{ts|\mathcal{I}}}\tilde{t}\\
\label{eq.40}&+\sqrt{\frac{2}{s}}\bigg(\frac{4\delta_{ts|\mathcal{I}}+2\sqrt{(t+(t-4)\delta_{ts|\mathcal{I}})\delta_{ts|\mathcal{I}}}}{t+(t-4)\delta_{ts|\mathcal{I}}}
+\sqrt{2}\bigg)\|x_{-\max(s)}\|_{2,\mathcal{I}}
\end{align}
for more details, see Appendix. Obviously observe that the upper bound of the error estimation given by (\ref{eq.39}) is much better than that determined by (\ref{eq.40}). In addition, even though set $d_i=1~(i=1,\cdots,l)$, we couldn't derive the general result coincided with \cite{Zhang and Li}. Consequently, the methodology of the proof that we employ for Theorem $1.1$ is preferable.
\end{remark}

\begin{corollary}
Under the same condition as in Theorem $1.1$, suppose that the noise term $\xi=0$ and the signal $x$ is block $s$-sparse, then $x$ can be perfectly recovered through the $l_2/l_1$ minimization (\ref{eq.7}).
\end{corollary}

\begin{remark}
The above result involving the noise-free and block $s$-sparse situation follows directly from Theorem $1.1$.
\end{remark}

The following result states that the bound of the block RIC $\delta_{ts|\mathcal{I}}<t/(4-t)$ with $0<t<4/3$, $ts\geq2$ is sharp for assured recovery in the noise-free setting.

\begin{theorem}
Suppose $s\geq1$ is an integer. If $\delta_{ts|\mathcal{I}}<t/(4-t)+\varepsilon$ with $0<t<4/3$ and $\varepsilon>0$, then the block $s$-sparse signal can not be exactly reconstructed via the $l_2/l_1$-minimization (\ref{eq.7}). Concretely, there is a sensing matrix $\Phi$ with $\delta_{ts|\mathcal{I}}=t/(4-t)$ and a block $s$-sparse $x_0$ satisfying $x^*\neq x_0$, where $x^*$ is the solution to (\ref{eq.7}).
\end{theorem}

The remainder of this article is organized as following. In Section $2$, we will provide some technical lemmas. In Section $3$, we will offer the proofs of main results. In Section $4,$ we draw a conclusion for this paper.

\section{Auxiliary lemmas}
\label{sec2}
All over this article, we utilize the notations below. Vector $x_{\Pi}$ indicates that it holds these blocks indexed by $\Pi$ of the vector $x$ and otherwise zero. For any block $s$-sparse vector, $\|x\|_{2,\infty}=\max_{1\leq i\leq s}\|x[i]\|_2.$ $\mbox{supp}(x)=\{i:\|x[i]\|_2\neq0\}$ denotes the block support of $x$.

The following two lemmas are necessary to the proof of the main results whose proofs are similar to that of Lemmas $1,~2$ \cite{Zhang and Li}. Denote $C^m_s=(^s_m).$
\begin{lemma}
Given vectors $\{x_i:~i\in \Pi\}$ in a vector space $X$ with inner product $<\cdot>,$ where $\Pi$ is an index set satisfying $|\Pi|=s.$ Suppose that we select all subsets $\Pi_i\in\Pi$ meeting $|\Pi_i|=m,$ $i\in\mathcal{J}$ with $|\mathcal{J}|=C^m_s,$ then
\begin{align}\label{eq.11}
\sum_{i\in\mathcal{J}}\sum_{j\in\Pi_i}x_j=C^{m-1}_{s-1}\sum_{i\in\Pi}v_i~(m\geq1),
\end{align}
and
\begin{align}\label{eq.12}
\sum_{i\in\mathcal{J}}\sum_{j\neq k\in\Pi_i}<x_j,x_k>=C^{m-2}_{s-2}\sum_{j\neq k\in\Pi}<x_j,x_k>~(m\geq2).
\end{align}
\end{lemma}

\begin{lemma}
Given a matrix $\Phi\in\mathbb{R}^{M\times N},$ we decompose $\Phi$ as a concatenation of column-blocks $\Phi[i]$ with size $M\times d_i$, say,
$$\Phi=[\underbrace{\phi_1,\cdots,\phi_{d_1}}_{\phi[1]},\underbrace{\phi_{d_1+1},\cdots,\phi_{d_1+d_2}}_{\phi[2]},\cdots,\underbrace{\phi_{N-d_l+1},\cdots,\phi_N}_{\phi[l]}],$$
and a vector $x\in\mathbb{R}^N~(l\geq2)$ that has the sparse pattern determined by (\ref{eq.38}) and put $\Omega=\{1,2,\cdots,l\}$. Suppose that we select all subsets $\Pi_i\subset\Omega$ fulfilling $|\Pi_i|=m,$ $i\in \mathcal{J}$ with $|\mathcal{J}|=C^m_l,$ and all subsets $\Lambda_j\subset\Omega$ obeying $|\Lambda_j|=n,~j\in\mathcal{K}$ with $\mathcal{K}=C^n_l.$ Then
\begin{align}\label{eq.13}
\sum_{i\in\mathcal{J}}\frac{(l-n)\|\Phi x_{\Pi_i}\|^2_2}{m|\mathcal{J}|}-\sum_{j\in\mathcal{K}}\frac{(l-m)\|\Phi x_{\Lambda_j}\|^2_2}{n|\mathcal{K}|}=\frac{(m-n)\|\Phi x\|^2_2}{l},
\end{align}
and as $l\geq m+n,$
\begin{align}\label{eq.14}
\sum_{\Pi_i\bigcap\Lambda_j=\emptyset}\frac{l-m-n}{mnl|\mathcal{J}|C^n_{l-m}}\bigg(\frac{mnl}{l-m-n}\|\Phi (x_{\Pi_i}+x_{\Lambda_j})\|^2_2-\|\Phi (nx_{\Pi_i}-mx_{\Lambda_j})\|^2_2\bigg)=\frac{(m+n)^2\|\Phi x\|^2_2}{l^2}.
\end{align}
\end{lemma}

The following lemma offers a crucial technical tool to the proof of our main theorems which is from \cite{Chen and Li}. For any vector $x$ with sparse structure defined by (\ref{eq.38}), $\|x\|_{2,2}=(\sum^l_{i=1}\|x[i]\|^2_2)^{1/2}$.
\begin{lemma}
For a positive number $\alpha$ and a positive integer $s$, the block polytope $\tau(\alpha,s)\subset\mathbb{R^N}$ is defined by
$$\tau(\alpha,s)=\{x\in\mathbb{R}^N:~\|x\|_{2,\infty}\leq\alpha,\|x\|_{2,\mathcal{I}}\leq s\alpha\}.$$
For any $x\in\mathbb{R}^N$, the set of block sparse vectors $\mathcal{U}(\alpha,s,x)\subset\mathbb{R}^N$ is defined by
$$\mathcal{U}(\alpha,s,x)=\{u\in\mathbb{R}^N:\mbox{supp}(u)\subseteq \mbox{supp}(x),\|u\|_{2,0}\leq s,\|u\|_{2,\mathcal{I}}=\|x\|_{2,\mathcal{I}},\|u\|_{2,\infty}\leq\alpha\}.$$
Then we can represent any $x\in\tau(\alpha,s)$ as
$$x=\sum_i\lambda_iu_i,$$
where $u_i\in\mathcal{U}(\alpha,s,x)$, $0\leq\lambda_i\leq1$, $\sum_i\lambda_i=1$, and $\sum_i\lambda_i\|u_i\|^2_{2,2}\leq s\alpha^2.$
\end{lemma}

The following lemma is necessary to the proof of the main results, whose proof is similar to that of Lemma $4.1$ \cite{Cai and Zhang}. Here we omit the detail.
\begin{lemma}
Let $\kappa\geq2$, $s\geq2$. For all measurement matrixes $\Phi$ with the size $M\times N~(M<N)$,  we obtain $\delta_{\kappa s|\mathcal{I}}\leq(2\kappa-1)\delta_{s|\mathcal{I}}$.
\end{lemma}

\section{Proofs of main results}

\noindent \textbf{Proof of Theorem $1.1$.}
First, suppose that $ts$ is an integer. As aforementioned, $h_{\max(s)}$ stands for $x$ keeping all but the $s$ largest blocks in $l_2$ norm set to zero. Let $h_{\max(s)}=h-h_{-\max(s)}$, $x^*=x+h$. Similar to the proof of Lemma $3.1$ \cite{Lin and Li}
, we could deduce
\begin{align}\label{eq.15}
\|h_{-\max(s)}\|_{2,\mathcal{I}}\leq\|h_{\max(s)}\|_{2,\mathcal{I}}+2\|x_{-\max(s)}\|_{2,\mathcal{I}}.
\end{align}
Select positive integers $m$ and $n$ satisfying $n\leq m\leq s$ and $m+n=st$. Subsets $\Pi_i,\Lambda_j\subset\{1,2,\cdots,s\}$ represent all the possible index set that $|\Pi_i|=m,|\Lambda_j|=n$ with $i\in\mathcal{J}$ and $j\in\mathcal{K}$ that $|\mathcal{J}|=C^m_s$ and $|\mathcal{K}|=C^n_s~(C^n_s=(^s_n))$.

Denote $$r=\frac{\|h_{\max(s)}\|_{2,\mathcal{I}}+2\|x_{-\max(s)}\|_{2,\mathcal{I}}}{s}.$$
Due to
\begin{align}
\notag\|h_{-\max(s)}\|_{2,\mathcal{I}}\leq sr=n\frac{s}{n}r,
\end{align}
and
\begin{align}\label{eq.16}
\notag\|h_{-\max(s)}\|_{2,\infty}&\leq\frac{\|h_{\max(s)}\|_{2,\mathcal{I}}}{s}\\
\notag&\leq\frac{\|h_{\max(s)}\|_{2,\mathcal{I}}+2\|x_{-\max(s)}\|_{2,\mathcal{I}}}{s}\\
&\leq r\leq\frac{s}{n}r.
\end{align}
Making use of (\ref{eq.16}) and Lemma $2.3$, we have $h_{-\max(s)}=\sum_i\lambda_iu_i$, where $u_i$ is block $n$-sparse, $0\leq\lambda_i\leq1$ with $\sum_i\lambda_i=1$, and $\mbox{supp}(u_i)\subset \mbox{supp}\left(h_{-\max(s)}\right)$, $\|u_i\|_{2,\mathcal{I}}=\|h_{-\max(s)}\|_{2,\mathcal{I}}$, $\|u_i\|_{2,\infty}\leq sr/n$, and
\begin{align}\label{eq.17}
\sum_i\lambda_i\|u_i\|^2_{2,2}\leq n\left(\frac{s}{n}r\right)^2=\frac{s^2r^2}{n}.
\end{align}
Analogously, we can decompose $h_{-\max(s)}$ as
\begin{align}
\notag h_{-\max(s)}=\sum_i\gamma_iv_i,\\
\notag h_{-\max(s)}=\sum_i\nu_iw_i,
\end{align}
where $v_i$ is block $m$-sparse, $w_i$ is block $(t-1)s$-sparse with
\begin{align}
\label{eq.18}\sum_i\gamma_i\|v_i\|^2_{2,2}\leq\frac{s^2r^2}{m}\\
\label{eq.19}\sum_i\nu_i\|w_i\|^2_{2,2}\leq\frac{sr^2}{t-1}.
\end{align}
Notice that $h_{\max(s)}$ is block $s$-sparse, and utilizing Cauchy-Schwarz inequality to any block $s$-sparse vector $x$, $\|x\|^2_{2,\mathcal{I}}=(\sum_i\|x[i]\|_2\cdot1)^2\leq s\sum_i\|x[i]\|^2_2=s\|x\|^2_{2,2}$, we have
\begin{align}
\notag r^2&=s^{-2}\left(\|h_{\max(s)}\|_{2,\mathcal{I}}+2\|x_{-\max(s)}\|_{2,\mathcal{I}}\right)^2\\\
\notag&=s^{-2}\left(\|h_{\max(s)}\|^2_{2,\mathcal{I}}+4\|h_{\max(s)}\|_{2,\mathcal{I}}\|x_{-\max(s)}\|_{2,\mathcal{I}}+4\|x_{-\max(s)}\|^2_{2,\mathcal{I}}\right)\\
\notag&\leq s^{-2}\left(s\|h_{\max(s)}\|^2_{2,2}+4\sqrt{s}\|h_{\max(s)}\|_{2,2}\|x_{-\max(s)}\|_{2,\mathcal{I}}+4\|x_{-\max(s)}\|^2_{2,\mathcal{I}}\right)\\
\label{eq.20}&=s^{-1}\|h_{\max(s)}\|^2_{2,2}+4s^{-\frac{3}{2}}\|h_{\max(s)}\|_{2,2}\|x_{-\max(s)}\|_{2,\mathcal{I}}+4s^{-2}\|x_{-\max(s)}\|^2_{2,\mathcal{I}}.
\end{align}
For $1\leq t<4/3$, exploiting the notion and monotonicity (Page 1404 \cite{Lin and Li}) of $\delta_{s|\mathcal{I}}$, we have
\begin{align}
\notag\left<\Phi h_{\max(s)},\Phi h\right>&\leq\|\Phi h_{\max(s)}\|_2\|\Phi h\|_2\\
\notag&\leq\sqrt{1+\delta_{s|\mathcal{I}}}\|h_{\max(s)}\|_2\|\Phi h\|_2\\
\label{eq.21}&\leq\sqrt{1+\delta_{ts|\mathcal{I}}}\|h_{\max(s)}\|_2\|\Phi h\|_2.
\end{align}
Since $x^*$ is the feasible solve to (\ref{eq.9}), we have
\begin{align}
\label{eq.22}\|\Phi h\|_2\leq\|\Phi(x-x^*)\|_2\leq\|\Phi x-b\|_2+\|\Phi x^*-b\|_2\leq2\rho.
\end{align}
Putting (\ref{eq.22}) into (\ref{eq.21}), we have
\begin{align}
\label{eq.23}\left<\Phi h_{\max(s)},\Phi h\right>\leq2\rho\sqrt{1+\delta_{ts|\mathcal{I}}}\|h_{\max(s)}\|_2.
\end{align}
For simplicity, we use $\mathcal{G}_{m,n}$ for
\begin{align}
\notag \mathcal{G}_{m,n}:=&\frac{s-n}{mC^m_s}\sum_{i\in\mathcal{J},k}\lambda_k\left(m^2\|\Phi(h_{\Pi_i}+\frac{n}{s}u_k)\|^2_2-n^2\|\Phi(h_{\Pi_i}-\frac{m}{s}u_k)\|^2_2\right)\\
\label{eq.24}&+\frac{s-m}{nC^n_s}\sum_{j\in\mathcal{K},k}\gamma_k\left(n^2\|\Phi(h_{\Lambda_j}+\frac{m}{s}v_k)\|^2_2-m^2\|\Phi(h_{\Lambda_j}-\frac{n}{s}v_k)\|^2_2\right).
\end{align}
Let $\theta(m,n,t)=2mn(t-2)+(m-n)^2.$ The following two equalities both hold, whose proof that may use Lemma $2.2$ are similar to that of identity (14) and (15) \cite{Zhang and Li}. The detail process is omitted. The equality
\begin{align}
\notag \frac{\theta(m,n,t)(t-1)}{mnC^m_sC^n_{s-m}}&\sum_{\Pi_i\bigcap\Lambda_j=\phi}\left(\frac{mn}{t-1}\|\Phi(h_{\Pi_i}+h_{\Lambda_j})\|^2_2+\|\Phi(nh_{\Pi_i}-mh_{\Lambda_j})\|^2_2\right)\\
\label{eq.25}&=t\mathcal{G}_{m,n}+2mn(t-2)t^2\left<\Phi h_{\max(s)},\Phi h\right>
\end{align}
holds for $0<t<1.$ The equality
\begin{align}
\notag \theta(m,n,t)&\sum_k\nu_k\left(\|\Phi(h_{\max(s)}+(t-1)w_k)\|^2_2-\|(t-1)\Phi(h_{\max(s)}-w_k)\|^2_2\right)\\
\label{eq.26}&=-(3t-4)\mathcal{G}_{m,n}+2((t-1)s^2-mn)t^3\left<\Phi h_{\max(s)},\Phi h\right>
\end{align}
holds for $1\leq t<4/3.$  For notational convenience, set
\begin{align}
\notag \Delta_1&=\theta(m,n,t)\sum_k\nu_k\left(\|\Phi(h_{\max(s)}+(t-1)w_k)\|^2_2-\|(t-1)\Phi(h_{\max(s)}-w_k)\|^2_2\right),\\
\notag \Delta_2&=-(3t-4)\mathcal{G}_{m,n}+2((t-1)s^2-mn)t^3\left<\Phi h_{\max(s)},\Phi h\right>,\\
\notag \Delta_3&=\frac{\theta(m,n,t)(t-1)}{mnC^m_sC^n_{s-m}}\sum_{\Pi_i\bigcap\Lambda_j=\phi}\left(\frac{mn}{t-1}\|\Phi(h_{\Pi_i}+h_{\Lambda_j})\|^2_2+\|\Phi(nh_{\Pi_i}-mh_{\Lambda_j})\|^2_2\right),\\
\notag \Delta_4&=t\mathcal{G}_{m,n}+2mn(t-2)t^2\left<\Phi h_{\max(s)},\Phi h\right>.
\end{align}
As to $\theta(m,n,t)$, as $ts$ is even, we can set $m=n=ts/2$; as $ts$ is odd, we can put $m=n+1=(ts+1)/2$. It is no difficult to check that $\theta(m,n,t)<0$ for the both situations.

By exploiting the definition of order $tk$ block RIC and observing that $h_{\Pi_i},~v_i$ are block $m$-sparse and $h_{\Lambda_j},u_i$ are block $n$-sparse obeying $m+n=ts$, we have
\begin{align}
\notag \mathcal{G}_{m,n}\geq&\frac{s-n}{mC^m_s}\sum_{i\in\mathcal{J},k}\lambda_k\bigg(m^2(1-\delta_{ts|\mathcal{I}})\|h_{\Pi_i}+\frac{n}{s}u_k\|^2_2
-n^2(1+\delta_{ts|\mathcal{I}})\|h_{\Pi_i}-\frac{m}{s}u_k\|^2_2\bigg)\\
\notag&+\frac{s-m}{nC^n_s}\sum_{j\in\mathcal{K},k}\gamma_k\bigg(n^2(1-\delta_{ts|\mathcal{I}})\|h_{\Lambda_j}+\frac{m}{s}v_k\|^2_2-m^2(1+\delta_{ts|\mathcal{I}})\|h_{\Lambda_j}-\frac{n}{s}v_k\|^2_2\bigg).
\end{align}
Note that $<h_{\Pi_i},u_k>=<h_{\Lambda_j},v_k>=0$, because of the support of $h_{\Pi_i}~(h_{\Lambda_j})$ does not intersect with the support of $u_k~(v_k)$. Therefore,
\begin{align}
\notag \mathcal{G}_{m,n}\geq&\frac{s-n}{mC^m_s}\sum_{i\in\mathcal{J},k}\lambda_k\bigg(m^2(1-\delta_{ts|\mathcal{I}})\|h_{\Pi_i}\|^2_2
+\frac{m^2n^2}{s^2}(1-\delta_{ts|\mathcal{I}})\|u_k\|^2_2\\
\notag&-n^2(1+\delta_{ts|\mathcal{I}})\|h_{\Pi_i}\|^2_2-\frac{m^2n^2}{s^2}(1+\delta_{ts|\mathcal{I}})\|u_k\|^2_2\bigg)\\
\notag&+\frac{s-m}{nC^n_s}\sum_{j\in\mathcal{K},k}\gamma_k\bigg(n^2(1-\delta_{ts|\mathcal{I}})\|h_{\Lambda_j}\|^2_2
+\frac{m^2n^2}{s^2}(1-\delta_{ts|\mathcal{I}})\|v_k\|^2_2\\
\notag&-m^2(1+\delta_{ts|\mathcal{I}})\|h_{\Lambda_j}\|^2_2-\frac{m^2n^2}{s^2}(1+\delta_{ts|\mathcal{I}})\|v_k\|^2_2\bigg)\\
\notag&=\frac{s-n}{mC^m_s}((m^2-n^2)-(m^2+n^2)\delta_{ts|\mathcal{I}})\sum_{i\in\mathcal{J}}\|h_{\Pi_i}\|^2_2
-\frac{s-n}{mC^m_s}2\frac{m^2n^2}{s^2}C^m_s\delta_{ts|\mathcal{I}}\sum_{k}\lambda_k\|u_k\|^2_2\\
\notag&+\frac{s-m}{nC^n_s}(-(m^2-n^2)-(m^2+n^2)\delta_{ts|\mathcal{I}})\sum_{j\in\mathcal{K}}\|h_{\Lambda_j}\|^2_2
-\frac{s-m}{nC^n_s}2\frac{m^2n^2}{s^2}C^n_s\delta_{ts|\mathcal{I}}\sum_{k}\gamma_k\|v_k\|^2_2.
\end{align}
By using (\ref{eq.11}), we have
\begin{align}
\notag \mathcal{G}_{m,n}\geq&((m^2-n^2)-(m^2+n^2)\delta_{ts|\mathcal{I}})\frac{s-n}{mC^m_s}C^{m-1}_{s-1}\|h_{\max(s)}\|^2_2-\frac{2(s-n)mn^2}{s^2}\delta_{ts|\mathcal{I}}\sum_{k}\lambda_k\|u_k\|^2_2\\
\notag&+(-(m^2-n^2)-(m^2+n^2)\delta_{ts|\mathcal{I}})\frac{s-m}{nC^n_s}C^{n-1}_{s-1}\|h_{\max(s)}\|^2_2-\frac{2(s-m)m^2n}{s^2}\delta_{ts|\mathcal{I}}\sum_{k}\gamma_k\|v_k\|^2_2.
\end{align}
Obviously, for any vector $x\in\mathbb{R}^N$ with sparsity pattern determined by (\ref{eq.38}), we could rewrite $l_2$-norm $\|x\|_2$ as
$$\|x\|_2=\left(\sum^N_{i=1}x^2_i\right)^{\frac{1}{2}}=\left(\sum^l_{i=1}\|x[i]\|^2_2\right)^{\frac{1}{2}}=\|x\|_{2,2}.$$ Due to (\ref{eq.17}) and (\ref{eq.18}), we have
\begin{align}
\notag \mathcal{G}_{m,n}\geq&((m^2-n^2)-(m^2+n^2)\delta_{ts|\mathcal{I}})\frac{s-n}{s}\|h_{\max(s)}\|^2_2-\frac{2(s-n)mn^2}{s^2}\delta_{ts|\mathcal{I}}\frac{s^2r^2}{n}\\
\notag&+(-(m^2-n^2)-(m^2+n^2)\delta_{ts|\mathcal{I}})\frac{s-m}{s}\|h_{\max(s)}\|^2_2-\frac{2(s-m)m^2n}{s^2}\delta_{ts|\mathcal{I}}\frac{s^2r^2}{m}\\
\notag=&\bigg(\frac{(m^2-n^2)(m-n)}{s}-\frac{(m^2+n^2)(2s-(m+n))\delta_{ts|\mathcal{I}}}{s}\bigg)\|h_{\max(s)}\|^2_2\\
\notag&-2mn\delta_{ts|\mathcal{I}}r^2(2s-(m+n))\\
\label{eq.27}=&\left(t(m-n)^2+(m^2+n^2)(t-2)\delta_{ts|\mathcal{I}}\right)\|h_{\max(s)}\|^2_2+2mn\delta_{ts|\mathcal{I}}r^2s(t-2).
\end{align}

First, we consider the case of $1\leq t<4/3.$

Since $\theta(m,n,t)$ is not lager than $0$, $h_{max(s)}$ is block $s$-sparse and $w_k$ is block $(t-1)s$-sparse combining with the definition of  order $ts$ block RIC $\delta_{ts|\mathcal{I}}$, then
\begin{align}
\notag \Delta_1&\leq\theta(m,n,t)\sum_k\nu_k\bigg((1-\delta_{ts|\mathcal{I}})\|h_{\max(s)}+(t-1)w_k\|^2_2-(1+\delta_{ts|\mathcal{I}})\|(t-1)(h_{\max(s)}-w_k)\|^2_2\bigg).
\end{align}
Observe that the support of $h_{\max(s)}$ does not intersect with that of $w_k$, thus
\begin{align}
\notag \Delta_1 &\leq\theta(m,n,t)\sum_k\nu_k\bigg((1-\delta_{ts|\mathcal{I}})\|h_{\max(s)}\|^2_2+(1-\delta_{ts|\mathcal{I}})(t-1)^2\|w_k)\|^2_2\\
\notag &-(t-1)^2(1+\delta_{ts|\mathcal{I}})\|h_{\max(s)}\|^2_2-(t-1)^2(1+\delta_{ts|\mathcal{I}})\|w_k\|^2_2\bigg)\\
\notag &=\theta(m,n,t)\sum_k\nu_k\bigg(((1-(t-1)^2)-(1+(t-1)^2)\delta_{ts|\mathcal{I}})\|h_{\max(s)}\|^2_2\\
\notag &-2(t-1)^2\delta_{ts|\mathcal{I}}\|w_k\|^2_2\bigg).
\end{align}
By applying (\ref{eq.19}) to the above inequality, we have
\begin{align}
\Delta_1 &\leq\theta(m,n,t)
\bigg(((1-(t-1)^2)-(1+(t-1)^2)\delta_{ts|\mathcal{I}})\|h_{\max(s)}\|^2_2\\
\label{eq.28} &-2\delta_{ts|\mathcal{I}}sr^2(t-1)\bigg).
\end{align}
It follows from the assumption of Theorem $1.1$ that $s\geq3/2$, so it is no hard to see that
\begin{align}
\label{eq.29}mn\geq\frac{t^2s^2-1}{4}=\frac{(2-t)^2s^2-1}{4}+(t-1)s^2>(t-1)s^2,
\end{align}
for $1\leq t<4/3.$ Combining with (\ref{eq.23}), (\ref{eq.27}) and (\ref{eq.29}), we have
\begin{align}
\notag\Delta_2&\geq-(3t-4)\bigg(\left(t(m-n)^2+(m^2+n^2)(t-2)\delta_{ts|\mathcal{I}}\right)\|h_{\max(s)}\|^2_2+2mn\delta_{ts|\mathcal{I}}r^2s(t-2)\bigg)\\
\label{eq.30}&+4\rho\sqrt{1+\delta_{ts|\mathcal{I}}}((t-1)s^2-mn)t^3\|h_{\max(s)}\|_2.
\end{align}
Let $\Delta_1$ minus $\Delta_2$, then
\begin{align}
\notag0\leq&\theta(m,n,t)
\bigg(((1-(t-1)^2)-(1+(t-1)^2)\delta_{ts|\mathcal{I}})\|h_{\max(s)}\|^2_2-2\delta_{ts|\mathcal{I}}sr^2(t-1)\bigg)\\
\notag&+(3t-4)\bigg(\left(t(m-n)^2+(m^2+n^2)(t-2)\delta_{ts|\mathcal{I}}\right)\|h_{-\max(s)}\|^2_2+2mn\delta_{ts|\mathcal{I}}r^2s(t-2)\bigg)\\
\notag&-4\rho\sqrt{1+\delta_{ts|\mathcal{I}}}((t-1)s^2-mn)t^3\|h_{\max(s)}\|_2.
\end{align}
By using (\ref{eq.20}) to the above inequality, the fact that for any vector $x$, $\|x\|_{2,2}=(\sum^l_{i=1}\|x[i]\|^2_2)^{\frac{1}{2}}=\|x\|_2$ and some elementary calculations, then
\begin{align}
\notag0\leq&2((t-1)s^2-mn)t^2\bigg((t+(t-4)\delta_{ts|\mathcal{I}})\|h_{\max(s)}\|^2_2\\
\label{eq.31}&-\bigg(\frac{4\delta_{ts|\mathcal{I}}\|x_{-\max(s)}\|_{2,\mathcal{I}}}{\sqrt{s}}+2\rho t\sqrt{1+\delta_{ts|\mathcal{I}}}\bigg)\|h_{\max(s)}\|_2
-\frac{4\delta_{ts|\mathcal{I}}\|x_{-\max(s)}\|^2_{2,\mathcal{I}}}{s}\bigg).
\end{align}

Next, we take into account the case of $0<t<1.$ Utilizing Lemma $2.4$, we have
\begin{align}
\notag\|\Phi h_{\max(s)}\|^2_2&\leq(1+\delta_{s|\mathcal{I}})\|h_{\max(s)}\|^2_2\\
\notag&=\left(1+\delta_{\frac{1}{t}ts|\mathcal{I}}\right)\|h_{\max(s)}\|^2_2\\
\notag&\leq\left(1+\left(\frac{2}{t}-1\right)\delta_{ts|\mathcal{I}}\right)\|h_{\max(s)}\|^2_2\\
\label{eq.32}&\leq\frac{(1+\delta_{ts|\mathcal{I}})\|h_{\max(s)}\|^2_2}{t}.
\end{align}
By (\ref{eq.22}) and (\ref{eq.32}), we have
\begin{align}
\notag\left<\Phi h_{\max(s)},\Phi h\right>&\leq\|\Phi h_{\max(s)}\|_2\|\Phi h\|_2\\
\label{eq.33}&\leq2\rho\frac{\sqrt{(1+\delta_{ts|\mathcal{I}})t}}{t}\|h_{\max(s)}\|_2.
\end{align}
By taking advantage of the concept of order $ts$ block RIC and (\ref{eq.11}), we have
\begin{align}
\notag \Delta_3&=\frac{\theta(m,n,t)(t-1)}{mnC^m_sC^n_{s-m}}\sum_{\Pi_i\bigcap\Lambda_j=\phi}\left(\frac{mn}{t-1}\|\Phi(h_{\Pi_i}+h_{\Lambda_j})\|^2_2+\|\Phi(nh_{\Pi_i}-mh_{\Lambda_j})\|^2_2\right)\\
\notag &\leq\frac{\theta(m,n,t)(t-1)}{mnC^m_sC^n_{s-m}}\sum_{\Pi_i\bigcap\Lambda_j=\phi}\left(\frac{mn}{t-1}(1-\delta_{ts|\mathcal{I}})\|h_{\Pi_i}+h_{\Lambda_j}\|^2_2+(1+\delta_{ts|\mathcal{I}})\|nh_{\Pi_i}-mh_{\Lambda_j}\|^2_2\right)\\
\notag &=\frac{\theta(m,n,t)(t-1)}{mnC^m_sC^n_{s-m}}\bigg(
\frac{mn}{t-1}(1-\delta_{ts|\mathcal{I}})\bigg( C^n_{s-m}\sum_{i\in\mathcal{J}}\|h_{\Pi_i}\|^2_2+C^m_{s-n}\sum_{j\in\mathcal{K}}\|h_{\Lambda_j}\|^2_2\bigg)\\
\notag&+(1+\delta_{ts|\mathcal{I}})\bigg( n^2C^n_{s-m}\sum_{i\in\mathcal{J}}\|h_{\Pi_i}\|^2_2+m^2C^m_{s-n}\sum_{j\in\mathcal{K}}\|h_{\Lambda_j}\|^2_2\bigg)\bigg)\\
\notag &=\frac{\theta(m,n,t)(t-1)}{mnC^m_sC^n_{s-m}}\bigg(
\frac{mn}{t-1}(1-\delta_{ts|\mathcal{I}})\bigg( C^n_{s-m}C^{m-1}_{s-1}\|h_{\max(s)}\|^2_2+C^m_{s-n}C^{n-1}_{s-1}\|h_{\max(s)}\|^2_2\bigg)\\
\notag&+(1+\delta_{ts|\mathcal{I}})\bigg( n^2C^n_{s-m}C^{m-1}_{s-1}\|h_{\max(s)}\|^2_2+m^2C^m_{s-n}C^{n-1}_{s-1}\|h_{\max(s)}\|^2_2\bigg)\bigg)\\
\notag &=\frac{\theta(m,n,t)(t-1)}{mnC^m_sC^n_{s-m}}\bigg(
\frac{mn}{t-1}(1-\delta_{ts|\mathcal{I}})\left(C^n_{s-m}C^{m-1}_{s-1}+C^m_{s-n}C^{n-1}_{s-1}\right)\|h_{\max(s)}\|^2_2\\
\notag&+(1+\delta_{ts|\mathcal{I}})\left(n^2C^n_{s-m}C^{m-1}_{s-1}+m^2C^m_{s-n}C^{n-1}_{s-1}\right)\|h_{\max(s)}\|^2_2\bigg)\\
\label{eq:34}&=\theta(m,n,t)(t+(t-2)\delta_{ts|\mathcal{I}})t\|h_{\max(s)}\|^2_2.
\end{align}
By combining (\ref{eq.27}) with (\ref{eq.33}), we have
\begin{align}
\notag\Delta_4&\geq t\bigg(\left(t(m-n)^2+(m^2+n^2)(t-2)\delta_{ts|\mathcal{I}}\right)\|h_{\max(s)}\|^2_2+2mn\delta_{ts|\mathcal{I}}r^2s(t-2)\bigg)\\
\notag&+4mn\rho\sqrt{1+\delta_{ts|\mathcal{I}}}(t-2)t\sqrt{t}\|h_{\max(s)}\|_2.
\end{align}
Let $\Delta_3$ minus $\Delta_4$, then
\begin{align}
\notag0&\leq\theta(m,n,t)(t+(t-2)\delta_{ts|\mathcal{I}})\|h_{\max(s)}\|^2_2\\
\notag&-t\bigg(\left(t(m-n)^2+(m^2+n^2)(t-2)\delta_{ts|\mathcal{I}}\right)\|h_{\max(s)}\|^2_2+2mn\delta_{ts|\mathcal{I}}r^2s(t-2)\bigg)\\
\notag&-4mn\rho\sqrt{1+\delta_{ts|\mathcal{I}}}(t-2)t\sqrt{t}\|h_{\max(s)}\|_2\\
\notag&\leq2t(t-2)mn\bigg((t+(t-4)\delta_{ts|\mathcal{I}})\|h_{\max(s)}\|^2_2\\
\label{eq.35}&-\bigg(\frac{4\delta_{ts|\mathcal{I}}\|x_{-\max(s)}\|_{2,\mathcal{I}}}{\sqrt{s}}+2\rho \sqrt{(1+\delta_{ts|\mathcal{I}})t}\bigg)\|h_{\max(s)}\|_2
-\frac{4\delta_{ts|\mathcal{I}}\|x_{-\max(s)}\|^2_{2,\mathcal{I}}}{s}\bigg).
\end{align}
By (\ref{eq.29}) and the assumption of $\delta_{ts|\mathcal{I}}<\frac{t}{4-t}$, it is easy to see that the above two inequalities given by (\ref{eq.31}) and (\ref{eq.35}) are second-order inequalities about $\|h_{\max(s)}\|_2$, where the quadratic coefficients are negative.

Consequently, through a straightforward calculation, we have
\begin{align}
\notag\|h_{\max(s)}\|_2\leq&\frac{\frac{4\delta_{ts|\mathcal{I}}\|x_{-\max(s)}\|_{2,\mathcal{I}}}{\sqrt{s}}+2\rho \sqrt{1+\delta_{ts|\mathcal{I}}}\tilde{t}}{2(t+(t-4)\delta_{ts|\mathcal{I}})}\\
\notag&+(2(t+(t-4)\delta_{ts|\mathcal{I}}))^{-1}\bigg(\bigg(\frac{4\delta_{ts|\mathcal{I}}\|x_{-\max(s)}\|_{2,\mathcal{I}}}{\sqrt{s}}+2\rho \sqrt{1+\delta_{ts|\mathcal{I}}}\tilde{t}\bigg)^2\\
\notag&+16(t+(t-4)\delta_{ts|\mathcal{I}})\frac{\delta_{ts|\mathcal{I}}}{s}\|x_{-\max(s)}\|_{2,\mathcal{I}}\bigg)^{\frac{1}{2}},
\end{align}
where $\tilde{t}=\max\{t,\sqrt{t}\}.$ Note the fact that for fixed $0<q\leq1$, any non-negative number $x,y$, $(x+y)^q\leq x^q+y^q$. Hence,
\begin{align}
\label{eq.36}\|h_{\max(s)}\|_2\leq\frac{2\rho\sqrt{1+\delta_{ts|\mathcal{I}}}\tilde{t}+2\left(\delta_{ts|\mathcal{I}}
+\sqrt{(t+(t-4)\delta_{ts|\mathcal{I}})\delta_{ts|\mathcal{I}}}\right)\|x_{-\max(s)}\|_{2,\mathcal{I}}/\sqrt{s}}{{t+(t-4)\delta_{ts|\mathcal{I}}}}.
\end{align}
Easily verify that for any block $s$-sparse vector $x$, $\|x\|_{2,2}=\left(\sum_{i}\|x[i]\|^2_2\right)^{1/2}\leq\sqrt{\|x[i]\|_{2,\infty}}\sqrt{\|x[i]\|_{2,\mathcal{I}}}.$ And employing Cauchy-Schwarz to any block $s$-sparse vector $x$, $\|x\|_{2,\mathcal{I}}=\sum_i\|x[i]\|_2\cdot1\leq s^{\frac{1}{2}}(\sum_i\|x[i]\|^2_2)^{\frac{1}{2}}=s^{\frac{1}{2}}\|x\|_{2,2}$ and combining with (\ref{eq.15}), we have
\begin{align}
\notag\|h_{-\max(s)}\|_{2,2}&\leq\sqrt{ s^{-1}\|h_{\max(s)}\|_{2,\mathcal{I}}}\sqrt{\|h_{\max(s)}\|_{2,\mathcal{I}}+2\|x_{-\max(s)}\|_{2,\mathcal{I}}}\\
\notag&\leq \sqrt{s^{-1}\|h_{\max(s)}\|^2_{2,\mathcal{I}}+2s^{-1}\|h_{\max(s)}\|_{2,\mathcal{I}}\|x_{-\max(s)}\|_{2,\mathcal{I}}}\\
\label{eq.37}&\leq \sqrt{\|h_{\max(s)}\|^2_{2,2}+2s^{-\frac{1}{2}}\|h_{\max(s)}\|_{2,2}\|x_{-\max(s)}\|_{2,\mathcal{I}}}.
\end{align}
By utilizing (\ref{eq.36}) and (\ref{eq.37}), we obtain
\begin{align}
\notag \|h\|_2&=(\|h_{\max(s)}\|^2_2+\|h_{-\max(s)}\|^2_2)^{\frac{1}{2}}\\
\notag&\leq\left(\|h_{\max(s)}\|^2_2+\|h_{\max(s)}\|^2_{2,2}+2s^{-\frac{1}{2}}\|h_{\max(s)}\|_{2,2}\|x_{-\max(s)}\|_{2,\mathcal{I}}\right)^{\frac{1}{2}}\\
\notag&=\left(2\|h_{\max(s)}\|^2_2+2s^{-\frac{1}{2}}\|h_{\max(s)}\|_{2,2}\|x_{-\max(s)}\|_{2,\mathcal{I}}\right)^{\frac{1}{2}}\\
\notag&\leq\left((\sqrt{2}\|h_{\max(s)}\|_2)^2+2\sqrt{2}\|h_{\max(s)}\|_{2,2}(2s)^{-\frac{1}{2}}\|x_{-\max(s)}\|_{2,\mathcal{I}}
+\left((2s)^{-\frac{1}{2}}\|x_{-\max(s)}\|_{2,\mathcal{I}}\right)^2\right)^{\frac{1}{2}}\\
\notag&=\sqrt{2}\|h_{\max(s)}\|_2+(2s)^{-\frac{1}{2}}\|x_{-\max(s)}\|_{2,\mathcal{I}}\\
\notag&\leq\frac{2\sqrt{2}\rho\sqrt{1+\delta_{ts|\mathcal{I}}}\tilde{t}+2\sqrt{2}(\delta_{ts|\mathcal{I}}
+\sqrt{(t+(t-4)\delta_{ts|\mathcal{I}})\delta_{ts|\mathcal{I}}})\|x_{-\max(s)}\|_{2,\mathcal{I}}/\sqrt{s}}{{t+(t-4)\delta_{ts|\mathcal{I}}}}\\
\notag&+\notag\|x_{-\max(s)}\|_{2,\mathcal{I}}/\sqrt{2s}\\
\notag&\leq \frac{2\sqrt{2}\rho\sqrt{1+\delta_{ts|\mathcal{I}}}\tilde{t}}{t+(t-4)\delta_{ts|\mathcal{I}}}\\
\notag&+\frac{1}{2}\sqrt{\frac{2}{s}}\bigg(\frac{4\left(2\delta_{ts|\mathcal{I}}+\sqrt{(t+(t-4)\delta_{ts|\mathcal{I}})\delta_{ts|\mathcal{I}}}\right)}{t+(t-4)\delta_{ts|\mathcal{I}}}
+1\bigg)\|x_{-\max(s)}\|_{2,\mathcal{I}}.
\end{align}

If $ts$ is not an integer, we set $t's=[ts],$ then $t's$ is an integer with $t'>t.$  For $t'<4/3$, we have $\delta_{t's|\mathcal{I}}=\delta_{ts|\mathcal{I}}<t/(4-t)<t'/(4-t')$. Analogous to the proof above, we can prove the result under the condition of $\delta_{ts|\mathcal{I}}<t/(4-t)$ with $ts\notin\mathbb{Z}.$

\qed

Now we prove Theorem $1.2.$

Denote
$$x_1=(2sd)^{-\frac{1}{2}}[\underbrace{\overbrace{1,\cdots,1}^d,\cdots,\overbrace{1,\cdots,1}^d}_{2s~\mbox{blocks}},0,\cdots,0]^T\in\mathbb{R}^N,$$
where $2s<l,$ $\mathcal{I}=\{d_1=d,d_2=d,\cdots,d_{2s}=d,d_{2s+1},\cdots,d_l\}$ and $\|x_1\|_2=1.$ The linear transformation $\Phi:\mathbb{R}^N\rightarrow\mathbb{R}^N$ is defined as follows:
$$\Phi x=\frac{1}{\sqrt{1-\frac{t}{4}}}(x-<x_1,x>x_1),$$
for any $x\in\mathbb{R}^N.$ For block $[ts]$-sparse signal $x\in\mathbb{R}^N$, then
\begin{align}
\notag \|\Phi x\|^2_2&=\frac{1}{1-\frac{t}{4}}\left(\| x\|^2_2-2<x_1,x>^2+<x_1,x>^2\| x_1\|^2_2\right)\\
\notag&=\frac{1}{1-\frac{t}{4}}\left(\| x\|^2_2-<x_1,x>^2\right).
\end{align}
Applying the Cauchy-Schwarz's inequality, we have
\begin{align}
\notag0&\leq |<x_1,x>|=|\left<x_1[\mbox{supp}(x)],x\right>|\\
\notag&\leq\|x_1[\mbox{supp}(x)]\|_2\|x\|_2\leq\|{x_1}_{\max([ts])}\|_2\|x\|_2\\
\notag&=\sqrt{\frac{[ts]}{2s}}\|x\|_2.
\end{align}
For $\varepsilon s>1$, then
\begin{align}
\notag \|\Phi x\|^2_2&\geq\frac{1}{1-\frac{t}{4}}\left(1-\frac{[ts]}{2s}\right)\| x\|^2_2\\
\notag &\geq\frac{1}{1-\frac{t}{4}}\left(1-\frac{ts+1}{2s}\right)\| x\|^2_2\\
\notag &\geq\frac{1}{1-\frac{t}{4}}\left(1-\frac{ts+\varepsilon s}{2s}\right)\| x\|^2_2\\
\notag &=\frac{1}{1-\frac{t}{4}}\left(1-\frac{t}{2}-\frac{\varepsilon }{2}\right)\| x\|^2_2\\
\notag &=\left(1-\frac{1}{\frac{4}{t}-1}-\frac{2\varepsilon}{4-t}\right)\| x\|^2_2\\
\notag &>\left(1-\frac{t}{4-t}-\varepsilon\right)\| x\|^2_2.
\end{align}
In the other direction, it is easy to see that
\begin{align}
\notag \|\Phi x\|^2_2&\leq\frac{4}{4-t}\| x\|^2_2=\left(1+\frac{t}{4-t}\right)\| x\|^2_2\\
\notag&\leq\left(1+\frac{t}{4-t}+\varepsilon\right)\| x\|^2_2.
\end{align}
Hence, we obtain $\delta_{ts|\mathcal{I}}=\delta_{[ts]|\mathcal{I}}<t/(4-t)+\varepsilon$.

At last, denote
$$x_0=[\underbrace{\overbrace{1,\cdots,1}^d,\cdots,\overbrace{1,\cdots,1}^d}_{s~\mbox{blocks}},\underbrace{\overbrace{0,\cdots,0}^d,\cdots,\overbrace{0,\cdots,0}^d}_{s~\mbox{blocks}},0,\cdots,0]^T\in\mathbb{R}^N,$$
$$\hat{x}=[\underbrace{\overbrace{0,\cdots,0}^d,\cdots,\overbrace{0,\cdots,0}^d}_{s~\mbox{blocks}},\underbrace{\overbrace{-1,\cdots,-1}^d,\cdots,\overbrace{-1,\cdots,-1}^d}_{s~\mbox{blocks}},0,\cdots,0]^T\in\mathbb{R}^N,$$
Easily check that $\|x_0\|_{2,\mathcal{I}}=\|\hat{x}\|_{2,\mathcal{I}}=s\sqrt{d},$ and $x_0,\hat{x}$ are block $s$-sparse, $x_1=(2sd)^{-1/2}(x_0-\hat{x}).$ Since $\Phi x_1=0$, then $\Phi x_0=\Phi \hat{x}=b$. It is no possible to recover vectors $x_0,\hat{x}$ only based on the known measurement matrix $\Phi$ and the observation vector $b$.

\qed

\section{Conclusions}
In the article, we investigate the block sparse signal recovery drawn from incomplete undetermined system of linear equations, whose non-zero coefficients are aligned into blocks, that is to say, they appear in blocks rather than arbitrarily disperse over all the vector. Based on block RIP, we derive a novel sufficient condition. Under the condition, we can assuredly recover all block sparse signals in the noise-free situation and robustly reconstruct signals that aren't exactly block sparse in the noisy situation by $l_2/l_1$-minimization methodology. Furthermore, we provide a special example to indicate that the sufficient condition we obtain is sharp. As byproduct, when $t=1$, the result enhances the bound of the block RIC $\delta_{s|\mathcal{I}}$ in \cite{Lin and Li}.

\noindent {\bf Appendix}\\
\noindent {\bf Lemma A.1}(Lemma 5.3 \cite{Cai and Zhang}) Assume that $s\leq l$, $a_1\geq a_2\geq\cdots\geq a_l\geq0$ obeys $\sum^s_{i=1}a_i\geq\sum^l_{i=s+1}a_i$, then we have
\begin{equation}\label{eq.41}
\sum^l_{i=s+1}a^{\alpha}_i\leq\sum^s_{i=1}a^{\alpha}_i
\end{equation}
for all $\alpha\geq1$. Generally, assume that $a_1\geq a_2\geq\cdots\geq a_l\geq0$, $\psi\geq0$ such that $\sum^s_{i=1}a_i\geq\sum^l_{i=s+1}a_i$ holds, then we have
\begin{equation}\label{eq.42}
\sum^l_{i=s+1}a^{\alpha}_i\leq s\left(\sqrt[\alpha]{\frac{\sum^s_{i=1}a^{\alpha}_i}{s}}+\frac{\psi}{s}\right)^{\alpha}
\end{equation}
for all $\alpha\geq1$.

\noindent \textbf{Proof of Remark $1.5$.}
Applying Lemma $A.1$ to (\ref{eq.15}), we have
\begin{align}
\notag\sum^l_{i=s+1}\|h[i]\|^2_2\leq s\left(\sqrt{\frac{\sum^s_{i=1}\|h[i]\|^2_2}{s}}+\frac{2\|x_{-\max(s)}\|_{2,\mathcal{I}}}{s}\right)^2,
\end{align}
i.e.,
\begin{align}
\notag\|h_{-\max(s)}\|_{2,2}\leq \|h_{\max(s)}\|_{2,2}+\frac{2\|x_{-\max(s)}\|_{2,\mathcal{I}}}{\sqrt{s}}.
\end{align}
Accordingly,
\begin{align}
\notag \|h\|_2&=(\|h_{\max(s)}\|^2_2+\|h_{-\max(s)}\|^2_2)^{\frac{1}{2}}\\
\notag&\leq\left(\|h_{\max(s)}\|^2_2+\left(\|h_{\max(s)}\|_{2,2}+\frac{2\|x_{-\max(s)}\|_{2,\mathcal{I}}}{\sqrt{s}}\right)^2\right)^{\frac{1}{2}}\\
\notag&\leq\sqrt{2}\|h_{\max(s)}\|_2+\frac{2\|x_{-\max(s)}\|_{2,\mathcal{I}}}{\sqrt{s}}\\
\notag&\leq \frac{2\sqrt{2}\rho\sqrt{1+\delta_{ts|\mathcal{I}}}}{t+(t-4)\delta_{ts|\mathcal{I}}}\tilde{t}\\
\notag&+\sqrt{\frac{2}{s}}\bigg(\frac{4\delta_{ts|\mathcal{I}}+2\sqrt{(t+(t-4)\delta_{ts|\mathcal{I}})\delta_{ts|\mathcal{I}}}}{t+(t-4)\delta_{ts|\mathcal{I}}}
+\sqrt{2}\bigg)\|x_{-\max(s)}\|_{2,\mathcal{I}}.
\end{align}

\qed

\noindent {\bf Acknowledgments}

This work was supported by Natural Science Foundation of China (No. 61673015, 61273020) and Fundamental Research Funds for the Central
Universities(No. XDJK2015A007).

\end{document}